\documentclass[aps,prc,preprint,tightenlines,groupedaddress,nofootinbib,
showpacs,preprintnumbers,amsmath,amssymb,superscriptaddress]{revtex4}
\usepackage{graphicx}
\usepackage{dcolumn}
\usepackage{bm}
\usepackage{amsmath}
\begin{document}

\title{Calculations of Recombination Rates for Cold $^4$He Atoms from Atom-dimer Phase Shifts and Determination of Universal Scaling Functions}
\author{J. R. Shepard}
\affiliation{Department of Physics, University of Colorado, 
	Boulder CO 80309}
\date{\today} 
\begin{abstract}
 Three-body recombination rates for cold $^4$He are calculated with a new method which exploits the simple relationship between the imaginary part of the atom-dimer elastic scattering phase shift and the $S$-matrix for recombination. The elastic phase shifts are computed above breakup threshold by solving a three-body Faddeev equation in momentum space with inputs based on a variety of modern atom-atom potentials. Recombination coefficients for the HFD-B3-FCII potential agree very well with the only previously published results. Since the elastic scattering and recombination processes for $^4$He are governed by ``Efimov physics", they depend on universal functions of a scaling variable. The newly computed recombination coefficients for potentials other than HFD-B3-FCII make it possible to determine these universal functions for the first time.

\end{abstract}
\maketitle 

\section{Introduction}
\label{Sec:Intro}

Much of the current interest in the properties of cold dilute atomic systems
arises from the possibility of examining exotic three-body effects which
are generically referred to as ``Efimov physics''. Such effects arise when 
the two-atom (dimer) scattering length $a$ is large compared to $\ell$ 
the range of the atom-atom potential which is typically of order the atomic size. 
Some remarkable ``universal'' features emerge in this limit (see, {\it e.g.}, refs. \cite{Braaten:2004rn,Braaten:2006vd} and references therein). For example, it is rigorously true that, as the dimer scattering length diverges and the binding energy goes to zero, the atom-dimer scattering length also diverges and the three-atom (trimer) system develops an infinite tower of bound states whose energies form a geometric sequence with zero as an accumulation point. In physical systems, departures from this limiting case scenario typically arise when, for example, $\ell/a$ is finite. However, when such departures are small, universal Efimov physics will still dominate.

The history of Efimov physics resembles to some extent that of Bose-Eistein Condensation (BEC)  in that the theory was well in hand long before experimenters were able to confirm its predictions. Indeed, it was only within the past year that the Innsbruck group~\cite{naegerl-2006} , exploiting the technique of Feshbach resonances in an ultracold gas of $^{133}$Cs atoms, saw evidence of Efimov physics in three-body recombination rates. Most of the elements of their analysis of the data are summarized in a recent review article by Braaten and Hammer~\cite{Braaten:2006vd}. Such an analysis, based on concepts of universality, requires some independent empirical input. A recent paper by Braaten, Kang and Platter~\cite{Braaten:2006qx} examines the universality constraints available from cold $^4$He and how these apply to the case of $^{133}$Cs. $^4$He is an especially interesting case because ({\it i}) it is relatively simple theoretically and ({\it ii}) nature has kindly provided a dimer scattering length which is large enough ($\sim$10 nm) that the criterion for Efimov physics to appear is generously satisfied. On the theoretical side, atom-atom potentials have been developed to a high degree of sophistication. Four widely used potentials are HFD-B~\cite{Aziz:1987}, HFD-B3-FCII~\cite{Aziz:1995}, LM2M2~\cite{Aziz:1991} and TTY~\cite{Tang:1995}. These have been employed in numerous ``ab initio'' calculations of dimer and trimer properties (see, {\it e.g.}, refs.~\cite{Esry:1999,Motovilov:1999iz,Roudnev:2002ab}). These results in turn have been used as input for the universality-based analyses mentioned above. On the experimental side, $^4$He dimers have been observed~\cite{Luo:1993,Schollkopf:1994} and the atom-atom bond length has been determined to be 5.2$\pm$0.4 nm from which an atom-atom scattering length of 10.4$^{+0.8}_{-1.8}$ nm has been deduced. (Scattering lengths  obtained using the four atom-atom potentials mentioned above are consistent with this range of values.) $^4$He trimers in their ground state have also been observed~\cite{Schollkopf:1994,Bruch:2002} and a recent paper~\cite{Bruhl:2005} reports a bond length of 1.1$^{+0.4}_{-0.5}$ nm which is consistent with theoretical estimates. This same paper indicates that the trimer excited state was not observed. The upper limit for the concentration of excited trimer states was determined to be appreciably less than theoretical estimates, a result which led the authors to entertain the possibility that such a state does not exist.

It is clear from the above discussion that three-body recombination rates are crucial in establishing the extent to which Efimov physics can be realized in the laboratory. To date, however, only a single ``ab initio'' calculation exists. It was performed by Suno {\it et al}~\cite{Suno:2002} for $^4$He using HFD-B3-FCII. As pointed out in Ref.~\cite{Braaten:2006qx}, this is unfortunate because it greatly inhibits the unversality-based analysis. In the present work, we present new results for the $^4$He three-body recombination coefficients. These quantities are obtained for the HFD-B, LM2M2 and TTY potentials as well as for HFD-B3-FCII where excellent agreement with Suno {\it et al.} is achieved.  As will be discussed below, a new calculational technique is employed which appears to be accurate and computationally efficient. Finally, we employ our new recombination coefficients to extract for the first time the universal functions, $h_1$ and $h_2$, introduced in Ref.~\cite{Braaten:2006qx} which govern low breakup energy recombination rates when Efimov physics is dominant.

\section{Calculation of Recombination Rates}
\label{sec:K3}

The rate equation for the density of a cold thermal atomic gas is given by~\cite{Suno:2002} 
\begin{equation}
\frac{dn_A}{dt}=-\frac{3 K_3}{6} n_A^3
\label{Ratezero}
\end{equation}
where $n_A$ is the atom number density, $K_3$ is the recombination coefficient and where we have assumed that all three atoms involved in the recombination leave the trap. We also have ignored the contribution from collision induced dissociation processes. For $^4$He atom-dimer scattering, the only inelastic channel corresponds to three-body breakup. Therefore the recombination $S$-matrix is directly related to the inelastic part of the atom-dimer $S$-matrix. It follows~\cite{Suno:2002} that
\begin{equation}
K_3=\sum_L (2L+1)\ K_3(L)=\sum_L(2L+1)\frac{192\pi^2}{\mu k^4}\ |S^{in}_L|^2
\label{Ratetwo}
\end{equation}
where $\mu=m_{He}/\sqrt{3}$ and $k$ depends on the energy, $E$,  in the breakup channel via $k=\sqrt{2\mu E}$. Also, $L$ is the angular momentum in the scattering channel and $S^{in}_L$ is the inelastic $S$-matrix element in that partial wave. Via the argument presented above, we have
\begin{equation}
|S^{in}_L|^2=1-{\rm e}^{-4\delta_I (L)}
\label{Smat}
\end{equation}
where $\delta_I (L)$ is the imaginary part of the phase shift for elastic atom-dimer scattering. Hence, calculation of $K_3(L)$ is equivalent to finding the atom-dimer elastic phase shift above the breakup threshold. We accomplish this by solving the momentum-space Faddeev equation for three-body scattering, to which we now turn.

\begin{table}
\begin{tabular}{|c|c|c|c|c|c|c|c|c|}
 \hline
Pot & $B_2$ (mK) & $a_2$ (nm) & $B_3^{(1)}$ (mK) & $H_0$ & $a_3$ (nm) 
	& $K_3^{(0)}(L=0)$ (cm$^6$/s) & ${\bar a}_2$ (nm) & ${\bar a}_2/a_{0*}$ \\
 \hline
HFD-B3-FCII & 1.58730  &  9.114  & 2.624 \cite{Esry:unpub} & 0.2835 & 12.017  & 7.06 $\times 10^{-28}$ & 8.738 & 1.1477 \\
 \hline
HFD-B            & 1.68541  &  8.856  & 2.74 \cite{Motovilov:1999iz}    & 0.2872 & 12.149  & 4.10 $\times 10^{-28}$ & 8.480 & 1.1222 \\
 \hline
         LM2M2 & 1.30348 & 10.018  & 2.28  \cite{Motovilov:1999iz}   & 0.2932   & 11.422  & 24.7 $\times 10^{-28}$ & 9.643 & 1.2367 \\
 \hline
                TTY & 1.30962 &  9.996   & 2.28  \cite{Motovilov:1999iz}   & 0.2855   & 11.548  & 22.7 $\times 10^{-28}$ & 9.620 & 1.2270 \\
\hline
\end{tabular}
\caption{$^4$He dimer and trimer properties. Trimer results are from FF calculations.  \hfill\break $K_3^{(0)}=K_3$ for $E_{breakup}\rightarrow 0$. See text for definition of quantities shown.}
\label{Table1}
\end{table}

\begin{figure}[ht]
\vspace{0.50in}
\includegraphics[width=6in,angle=0]{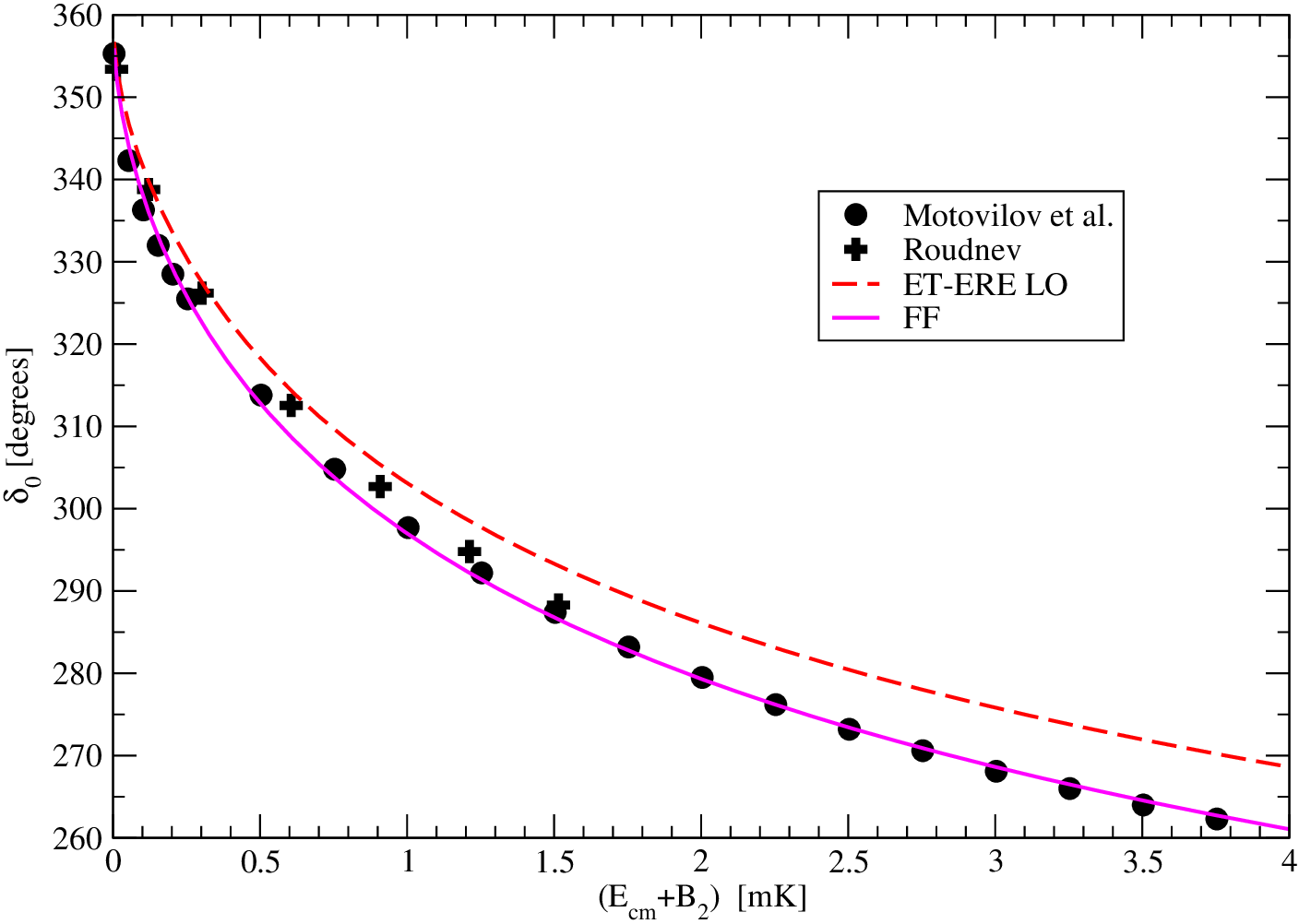}
\caption{(Color online).
         Atom-dimer phase shifts using potential LM2M2. Circles are from Motovilov {et al.}~\cite{Motovilov:1999iz}; crosses are from Roudnev~\cite{Roudnev:2002ab}. To be compared with Figure 3 of Ref.~\cite{Platter:2006ev}. Breakup threshold is at $E_{cm}+B_2=1.30348$ mK. ET-ERE NLO and NNLO calculations are virtually identical to the FF curve on this plot.}
\label{Fig1}
\end{figure}

Following, {\it e.g.}, Watson and Nuttall~\cite{Watson:1967}, we assume the $s$-wave atom-atom interaction is described by a separable potential: 
\begin{equation}
V(q,q')=g(q)\ \lambda\ g^*(q')
\label{Twobodypot}
\end{equation}
where $g(q)$ is a form factor (FF) to be discussed at length below. The atom-atom $t$-matrix is then
\begin{equation}
t(q,q';E^+)=g(q)\ \tau(E^+)\ g(q')
\label{TwobodyTmat}
\end{equation}
where $E^+=E+i\eta$, $\tau(E^+)=\lambda/[1-\lambda\ h(E^+)]$ and
\begin{equation}
h(E^+)=\int\frac{d^3k}{(2\pi)^3}\ \frac{|g(k)|^2}{E^+ - k^2/m}
\label{TwobodyhofE}
\end{equation}
where $m$ is the atomic mass. Following, {\it e.g.}, the standard Faddeev-based treatment of Ref.~\cite{Watson:1967}, we find that the fully off-shell atom-dimer $t$-matrix for three identical bosons in the $L$-th partial wave is 
\begin{equation}
T_L(p,k;E^+)=2{\cal Z}_L(p,k;E^+)+2\int\frac{d^3q}{(2\pi)^3}\ {\cal Z}_L(p,q;E^+)\ \tau\biggl(E^+-\frac{3q^2}{4m}\biggr)\ T_L(q,k;E^+)
\label{ThreebodyTmat}
\end{equation}
with
\begin{equation}
{\cal Z}_L(p,k;E^+)=\frac{1}{2}\int_{-1}^{+1}dx\ P_L(x)\ \frac{g(p/2+k)\ g^*(p+k/2)}{E^+-(p^2+k^2+pkx)/m} \label{Threebodyborn}
\end{equation}
where $P_L(x)$ is a Legendre polynomial. 
(The full derivation of Eq.~\ref{ThreebodyTmat} is somewhat lengthy. Ever longer is the full history leading to the development presented in Ref.~\cite{Watson:1967}. Neither the derivation nor the history can be fully recounted here but suffice it to say that the many references presented in Watson and Nuttall~\cite{Watson:1967} dating from the 1950's and 1960's contain the direct antecedents to, {\it e.g.}, the trimer bound state calculations of Lim {\it et al}.~\cite{Lim:1977}. The expressions for the three-body amplitudes contained in these early references have been re-derived many times since; 
Refs.~\cite{Bedaque:2002yg,Bedaque:2002mn,Bedaque:1998km,Bedaque:1998kg,Braaten:2001hf} are but a few representative examples. Note that, in virtually all of the these more modern papers, two-body {\it contact interactions} are assumed. See, however, Afnan and Phillips~\cite{Afnan:2003bs} for a modern reference in which expressions very similar to ours appear even though, in their work, the form factors $g(q)$ are introduced for pedagogical purposes only and are set to 1 early on.) The elastic scattering amplitude is then given by
\begin{equation}
f(\theta)=\sum_L(2L+1)\ f_L(p)\ P_L(\cos\theta)
\label{Threebodyfoftheta}
\end{equation}
where
\begin{equation}
f_L(p)=-\frac{m}{3\pi}\ Z^{1/2}\ T_L\bigl(p,p;\frac{3 p^2}{4m}-B_2+i\eta\bigr)=\frac{{\rm e}^{2i\delta_L(p)}-1}{2ip}
\label{Threebodyfofp}
\end{equation}
where the atom-dimer $t$-matrix is fully on-shell.
Here $B_2$ is the dimer binding energy which satisfies $1-\lambda h(-B_2)=0$ and 
the bound state normalization $Z$ is given by
\begin{equation}
Z^{-1/2}=-\frac{d h(E)}{dE}\bigg|_{E=-B_2}.
\label{Twobodynorm}
\end{equation}
If the form factor $g(q)$ is taken to be a $\delta$-function, then $h(E^+)=\frac{m \sqrt{-m E^+}}{4\pi}$ after the divergence is removed. Setting $4\pi/\lambda m=a_2$, the atom-atom scattering length, we find that Eq.~\ref{ThreebodyTmat} is the standard Effective Theory (ET) result at leading order (LO) in the Effective Range Expansion (ERE) of the atom-atom $t$-matrix. (See, {\it e.g.}, Refs.~\cite{Griesshammer:2004pe,Platter:2006ev} for extensive discussions of this treatment and for additional related references. We will subsequently refer to results based on this formulation as ET-ERE calculations.) Below the breakup threshold, the resulting Faddeev equation, Eq.~\ref{ThreebodyTmat}, can readily be solved as it stands by discretizing in momentum space. This yields a system of simultaneous linear equations which can be handled with standard linear algebra packages such as LAPACK. Above breakup, as noted long ago by Aaron and Amado~\cite{Aaron:1966} in the context of neutron-deuteron scattering, ${\cal Z}_L$ becomes ill-behaved along the real axis. A simple and efficient solution to this problem was given by Hetherington and Schick~\cite{Hetherington:1965} even longer ago. They suggested deforming the contour of integration over $q$ in Eq.~\ref{ThreebodyTmat} away from the real axis, solving the Faddeev equation for $T_L$ of complex argument, and then using the Faddeev equation once more to obtain $T_L\bigl(p,p;\frac{3 p^2}{4m}-B_2+i\eta\bigr)$. Finally, the relation given in Eq.~\ref{Threebodyfofp} is utilized to extract the phase shift, $\delta_L(p)$. When carefully applied, this prescription can yield accurate complex phase shifts above the breakup threshold and hence, via Eqs.~\ref{Ratetwo}\ and \ref{Smat}, numerical values for the recombination coefficient, $K_3$. It should be noted that care is required because the imaginary parts of the phase shifts are typically quite small. For $L=0$ near threshold, Im $\delta_L$ is typically of order $10^{-5}$ degree! The fact that we compute values of Im $\delta_L$ which are typically less than $10^{-8}$ degree just below threshold is an important necessary condition for trusting the small values we find above threshold. Another indication that our numerics are adequate in computing Im $\delta_L$ is that our values of $K_3(L=0)$ go to a constant as threshold is approached,  as required by general considerations~\cite{Nielsen:1999,Esry:2001}. This requires considerable numerical precision since, in this region, Im $\delta_0$ is proportional to $E^2$, where $E$ the total energy in the breakup channel.

\begin{table}
\begin{tabular}{|c|c|c|c|c|c|c|c|}
 \hline
Pot & Ref.~\cite{Motovilov:1999iz} & Ref.~\cite{Blume:2000} & Ref.~\cite{Roudnev:2002ab} & Ref.~\cite{Penkov:2003}  & Ref.~\cite{kolganova-2004-70}  & Ref.~\cite{Platter:2006ev} & Present work (FF)\\
\hline
HFD-B  & 13.5$\pm$0.5 &  -----  & 12.19$\pm$0.01 &    -----       & 12.47  &  -----  &  12.149 \\
\hline
LM2M2 & 13.1$\pm$0.5 &  12.6 & 11.54$\pm$0.01  & 11.425 & 11.87  &   ----- & 11.422 \\
\hline
TTY       &          -----            &  -----    & 11.58$\pm$0.01   &    -----     & 11.89   &   11.573 & 11.548 \\
\hline
\end{tabular}
\caption{$^4$He atom-dimer scattering lengths in nm compared with previous results.}
\label{Table2}
\end{table}

In the following section we present results using Eqs.~\ref{ThreebodyTmat} and \ref{Threebodyfofp}. In addition, we have done calculations at next-to-leading order (NLO) and next-to-next-to-leading order (NNLO) in the ET-ERE~\cite{Griesshammer:2004pe,Platter:2006ev}.  To facilitate comparison with the published results of the latter reference, we present our definitions of these calculations. Recall the effective range expansion of the two-body $s$-wave scattering amplitude:
\begin{equation}
p\cot\delta_0=-\frac{1}{a_2}+\frac{1}{2} r p^2 +{\cal O}[p^4].
\label{TwoBodyERE}
\end{equation}
where $a_2$ is the scattering length and $r$ is the range parameter. We note that for $^4$He atom-atom scattering, $a_2\simeq 10$ nm while $r\simeq0.75$ nm.  Then, at LO, the quantity $\tau(E^+-\frac{3q^2}{4m})$ in Eq.~\ref{ThreebodyTmat} may be expressed as
\begin{equation}
\tau\biggl(E^+-\frac{3q^2}{4m}\biggr)=-\frac{4\pi}{m} \frac{1}{-\gamma+\sqrt{\frac{3}{4}(q^2-k^2)+\gamma^2}}
\label{tauLO}
\end{equation}
where $\gamma=4\pi/\lambda m$ and where we have used $E^+=(\frac{3}{4}k^2-\gamma+i\eta)/m$. At LO, we take $\gamma=1/a_2$. At higher orders in the ET-ERE, we have
\begin{equation}
\tau\biggl(E^+-\frac{3q^2}{4m}\biggr)=-\frac{4\pi}{m} \frac{1}{-\gamma+\sqrt{\frac{3}{4}(q^2-k^2)+\gamma^2}}\biggl[1+\sum_{n=1}^{N_{max}}\biggl(\frac{\frac{r}{2}[\frac{3}{4}(q^2-k^2)+\gamma^2]}{-\gamma+\sqrt{\frac{3}{4}(q^2-k^2)+\gamma^2}}\biggr)^n \ \biggr]
\label{tauNNLO}
\end{equation}
where $N_{max}=1$ or 2 for NLO or NNLO, respectively. In this expression, we take
\begin{equation}
\gamma=\frac{1}{r}\biggl(1-\sqrt{1-\frac{2 r}{a_2}}\biggr)
\label{gamaNNLO}
\end{equation}
which means that $p=i\gamma$ gives the position of the bound state pole in the two-body $T$-matrix which corresponds to a zero in
\begin{equation}
	1/f_0(p)=p\cot\delta_0-ip
\label{B2NNLO}
\end{equation}
when terms of ${\cal O}[p^4]$ are dropped in the effective range expansion for $p\cot\delta_0$ (Eq.~\ref{TwoBodyERE}). Note that the expression for $\tau(E^+-\frac{3q^2}{4m})$ in Eq.~\ref{tauNNLO} ensures that the pole structure of that quantity is the same at NLO and NNLO as at LO~\cite{Griesshammer:2004pe}. 

In addition to ET-ERE calculations, we will also show results for which the form factor, {\it i.e.}, the $g(q)$ appearing in Eqs.~\ref{Twobodypot} and \ref{Threebodyborn}, is {\it not} assumed to be a delta function but rather is taken to be of the form
\begin{equation}
g(q)\rightarrow \biggl[1 + C\ \frac{q^2}{\beta^2}\biggr]\frac{1}{(1+q^2/\beta^2)^2}
\label{FormFactorTwo}
\end{equation}
where $\beta$ and $C$ are fixed by the low energy atom-atom phase shifts~\cite{Shepard:2007b}. We compute these phase shifts directly from the atom-atom potentials along with the corresponding dimer binding energies. 

The dimer binding energies and atom-atom scattering lengths for the four potentials considered here appear in Table \ref{Table1}. As discussed in detail in, {\it e.g.}, Refs. \cite{Bedaque:1998kg,Bedaque:1998km,Bedaque:1999ve,Bedaque:2002mn,Platter:2006ev}, the Faddeev equation for $L=0$ is a special case of considerable interest as it is here that Efimov physics is manifest. In the standard ET-ERE approach, $T_{L=0}$ at LO exhibits a log-periodic dependence on the upper limit, $\Lambda$, of the integration over $q$ in Eq.~\ref{ThreebodyTmat}. Moreover, a three-body force enters at lowest order in the calculation and is accounted for via
\begin{equation}
{\cal Z}_{L=0}\rightarrow{\cal Z}_{L=0}+\frac{2H_0}{\Lambda^2}
\label{BigH}
\end{equation}
where the three-body interaction strength $H_0$ is dimensionless. For the ET-ERE calculations reported here, we choose $\Lambda$ arbitrarily apart from the requirement that it to be large enough that the overall results do not depend on its magnitude. Then $H_0$ is adjusted to reproduce $B_3^{(1)}$, the binding energy of the trimer excited state which is known to be of an Efimov character~\cite{Lim:1977}. (Note that our trimer binding energies are defined so that zero corresponds to the three particles being at rest and far enough apart that they do not interact.) Consequently, the value of $H_0$ is $\Lambda$-dependent. When using the finite-range form-factors, the situation is somewhat different. Because the form-factors regulate the integrand in Eq.~\ref{ThreebodyTmat}, there is no log-periodic dependence on $\Lambda$. The three-body interaction must still be included, with its value determined as before. Now, however, $H_0$ is independent of the momentum cutoff.  A thorough discussion of form-factor calculations like these will be presented in a future publication~\cite{Shepard:2007b}. Values of $B_3^{(1)}$ and $H_0$ for form-factor calculations  (FF) are given in Table~\ref{Table1} for the four atom-atom potentials we consider.



\begin{figure}[ht]
\vspace{0.50in}
\includegraphics[width=6in,angle=0]{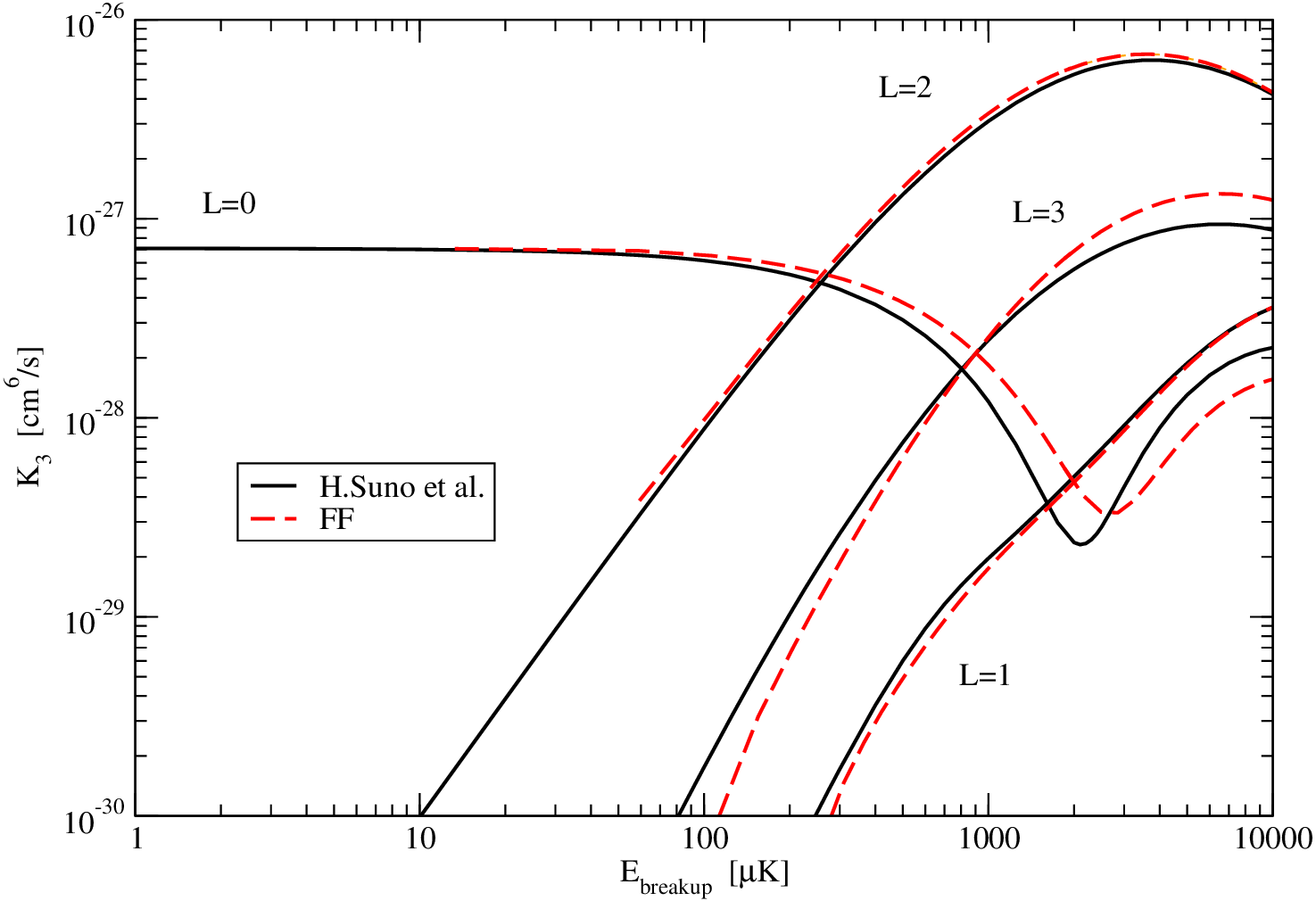}
\caption{(Color online).
         $K_3(L)$ computed using  the form factor $g(q)$ (FF) and atom-atom potential HFD-B3-FCII.
         Present results (dashed) are compared with those of Suno {\it et al.}~\cite{Suno:2002} (solid).
}
\label{Fig2}
\end{figure}


\section{Results}
\label{sec:results}

To begin this section, we focus on results for the atom-dimer scattering length. Table \ref{Table2} compares our FF values with those of previous works. Agreement with Roudnev~\cite{Roudnev:2002ab}, Pen'kov~\cite{Penkov:2003} and Platter and Phillips~\cite{Platter:2006ev} is quite good. The former calculations are based on differential Faddeev equations in configuration space in a total angular momentum representation. The author claims a two order of magnitude improvement in accuracy over earlier ``ab initio'' calculations of a similar nature by Motovilov {\it et al.}~\cite{Motovilov:1999iz}. Note that more recent results by the latter group appear in Table~\ref{Table2} under Ref.~\cite{kolganova-2004-70}. Platter and Phillips~\cite{Platter:2006ev} report a version of ET-ERE calculations using TTY. We find especially close agreement for LM2M2 with Pen'kov~\cite{Penkov:2003} whose work we became aware of after the present work was completed and whose method is similar in some respects to ours. We are able to confirm an observation by Pen'kov that $p \cot\delta_0$ for $^4$He atom-dimer scattering has an unusual low-energy behavior which greatly restricts the range  over which the low order effective range expansion is accurate. We also note that this behavior necessitates some care in finding the scattering length by, in effect, extrapolating $p \cot\delta_0$ to $p=0$. In particular, extrapolations which do not take into account this unusual behavior will yield scattering lengths which are too large. Such problems may account in part for some of the larger discrepancies with previous results appearing in Table \ref{Table2}. Figure~\ref{Fig1} compares our atom-dimer $L=0$ phase shifts for LM2M2 with those of Motovilov {\it et al.}~\cite{Motovilov:1999iz} and of Roudnev~\cite{Roudnev:2002ab}. We show results of our ET-ERE calculations at LO, NLO and NNLO along with form factor calculations (FF). (We remind the reader to look at Refs.~\cite{Griesshammer:2004pe,Platter:2006ev} for thorough discussions of ET-ERE calculations.) First we observe that there are significant differences between LO and higher order ET-ERE results but the ET-ERE calculations are well-converged at NLO. Moreover, the NLO and NNLO ET-ERE results agree well with the FF phase shifts. As will be discussed in a future publication~\cite{Shepard:2007b}, this behavior is readily explained. First, range effects are small here since $r/a_2$ for $^4$He atom-atom scattering ($a_2$ and $r$ are the two atom scattering length, respectively, in the effective range expansion of $p \cot\delta$) is roughly $0.75$ nm / $10$ nm. Also, the FF calculations include, by construction, certain of the finite range effects to {\it all} orders~\cite{Shepard:2007b}. Figure~\ref{Fig2} shows the present phase shifts lie consistently below those of Roudnev but are in good agreement with the calculations of Motovilov {\it et al}. All our calculated phase shifts (except for ET-ERE LO) are indistinguishable from those of Platter and Phillips~\cite{Platter:2006ev}. 
\begin{table}
\begin{tabular}{|c|c|c|c|c|}
 \hline
Pot & ET-ERE LO & ET-ERE NLO & ET-ERE NNLO & FF \\
\hline
HFD-B3-FCII & 9.833 (-18.2) &  12.029 (+0.10) & 12.024 (+0.06) & 12.017    \\
\hline
HFD-B            &   9.957 (-18.0) & 12.163 (+0.12) &  12.160 (+0.09) & 12.149    \\
\hline
LM2M2          &    9.269  (-18.8) &  11.446 (+0.21) & 11.435 (+0.12) & 11.422   \\
\hline
TTY                &    9.388  (-18.7) &  11.566 (+0.16) & 11.557 (+0.08) & 11.548  \\
\hline
\end{tabular}
\caption{$^4$He atom-dimer scattering lengths for various potentials and calculation types. Values in parentheses are percent deviations of ET-ERE calculations from the FF results.}
\label{Table3}
\end{table}
\begin{figure}[ht]
\vspace{0.50in}
\includegraphics[width=6in,angle=0]{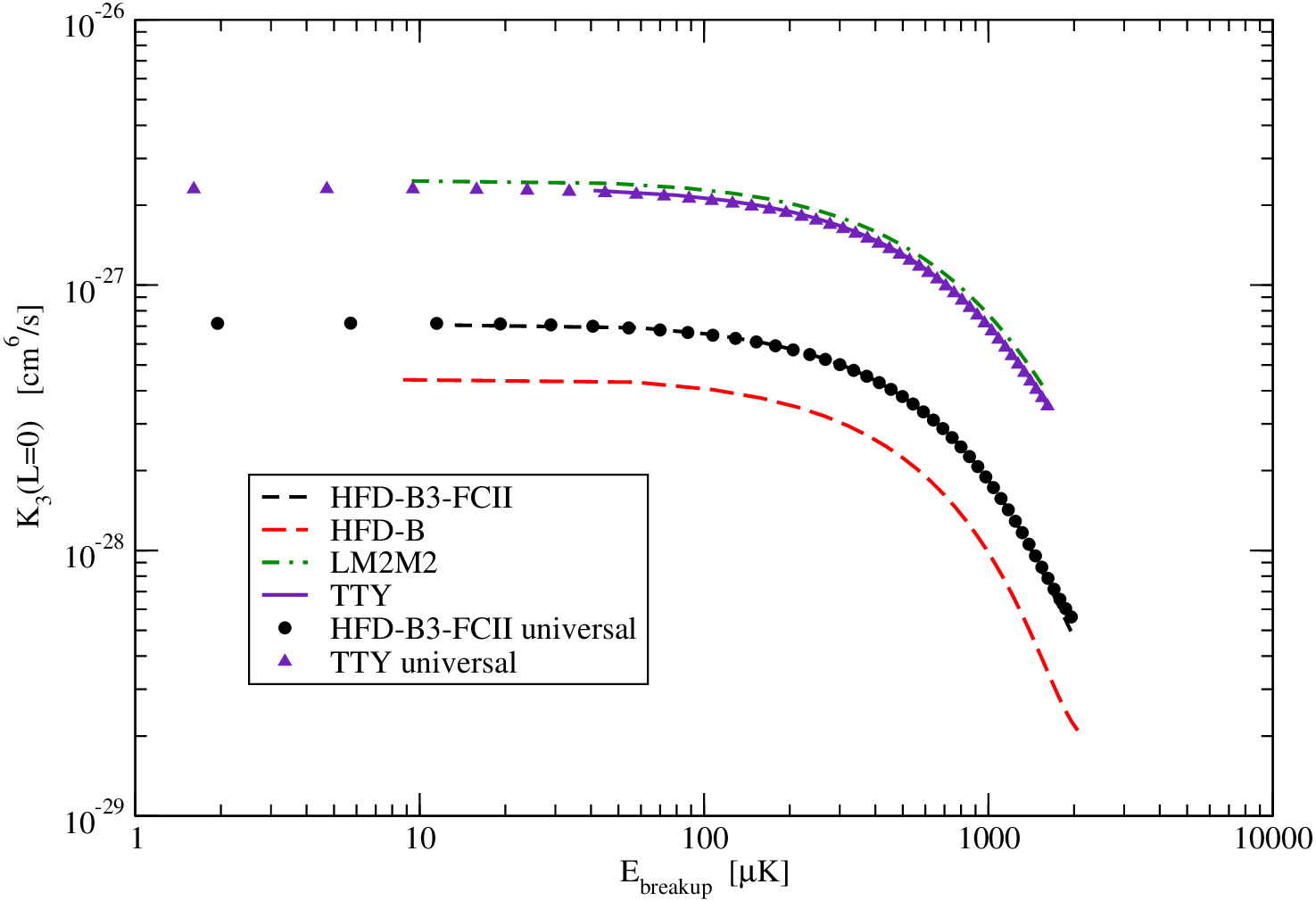}
\caption{(Color online).
         $K_3(L=0)$'s computed using the form factor $g(q)$ (FF) for various atom-atom potentials. The solid lines represent full calculations while the dashed curves are computed using the universal functions $h_1$ and $h_2$ and the values of ${\bar a}_2$ and $a_{0*}$ for potentials HFD-B3-FCII and TTY. Universal calculations for HFD-B and LM2M2 agree with their full counterparts by construction. See text for discussion.
}
\label{Fig3}
\end{figure}


Our values for scattering lengths computed with different atom-atom potentials and using different methods of calculation are displayed in Table \ref{Table3}. For the ET-ERE calculations at NLO and NNLO, our values depart from those of Platter and Phillips~\cite{Platter:2006ev} only at the level of +0.27 and -0.14~\%, respectively. Based on the small value of $r/a_2\simeq 0.075$, we expect the ET-ERE calculations to converge rapidly, an expectation which is readily confirmed by examination of Table~\ref{Table3}. Moreover, it appears that the ET-ERE calculations are consistently converging toward the FF results. Such behavior would be consistent with the assertion made above that the FF calculations include some range effects to all orders~\cite{Shepard:2007b}. Table~\ref{Table3} also makes it clear that there are significant differences between the HFD-B3-FCII and HFD-B atom-atom potentials on the one hand and the LM2M2 and TTY potentials on the other. Such differences will reappear in the discussions to follow.

\begin{figure}[ht]
\vspace{0.50in}
\includegraphics[width=6in,angle=0]{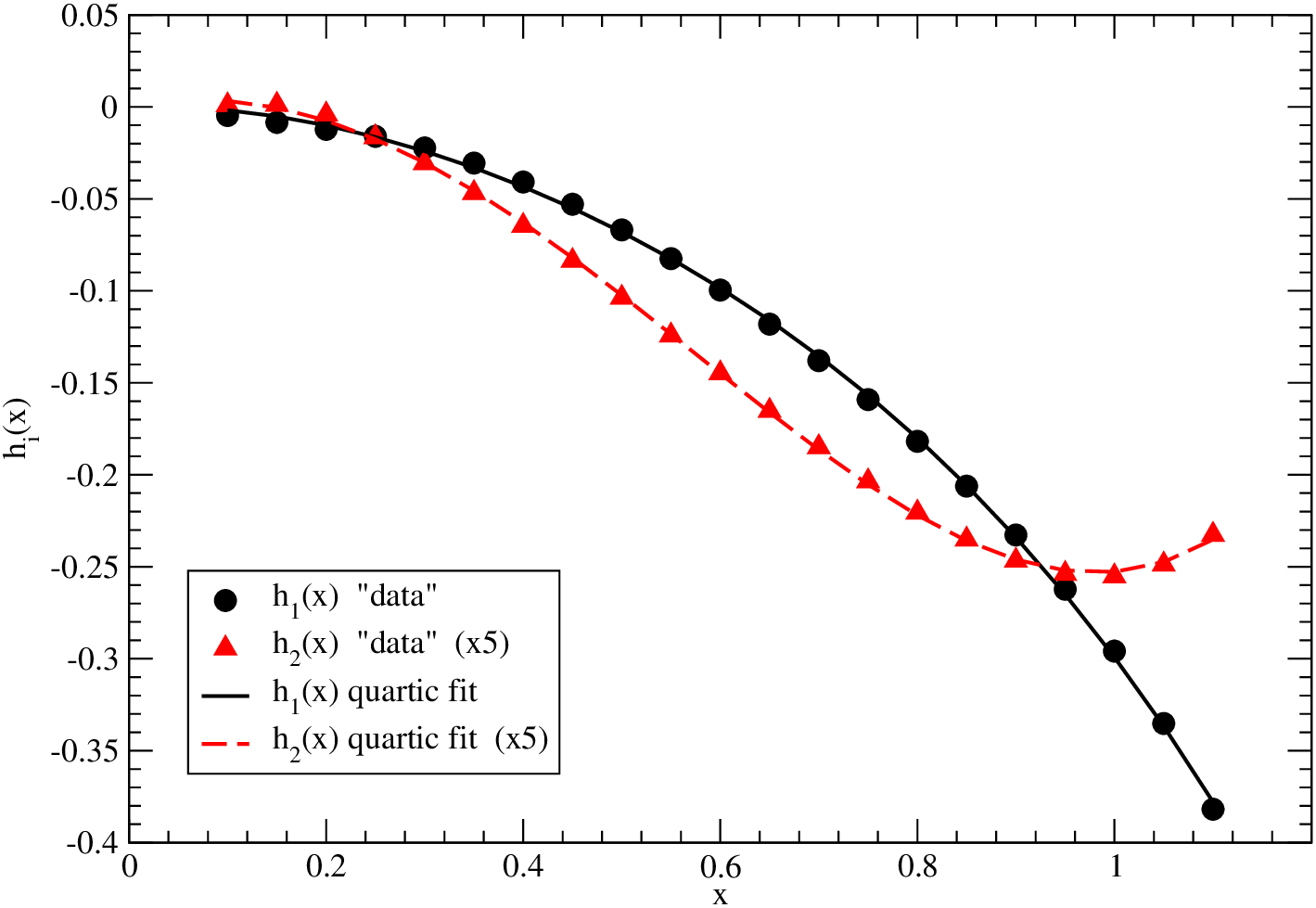}
\caption{(Color online).
         Universal functions $h_1(x)$ and $h_2(x)$ as defined in Ref.~\cite{Braaten:2006qx} and appearing here  in Eq.~\ref{astar}. The circles ($h_1$) and triangles  ($h_2$) are values extracted from HFD-B3-FCII and LM2M2 calculations of $K_3(L=0)$. The fits are: $h_1(x)=0.0179358\ x - 0.379918\ x^2 + 0.22438\ x^3 - 0.161562\ x^4$ and 
         $h_2(x)=0.0214188\ x - 0.152046\ x^2 + 0.029073\ x^3 + 0.051001\ x^4$. 
}
\label{Fig4}
\end{figure}


We now turn to evaluation of the $^4$He three-body recombination coefficients. FF results for $K_3(L=0)$ in the low energy limit are displayed in Table \ref{Table1}. Our value for HFD-B3-FCII of $7.06\ \times 10^{-28}$ cm$^6$/s is in striking agreement with the only previously published value of of this quantity, namely $7.09\ \times 10^{-28}$ cm$^6$/s as reported by Suno, Esry, Greene and Burke~\cite{Suno:2002}. Figure \ref{Fig2} compares our results for $K_3(L)$, $L\leq 3$ as a function of breakup energy with those of Suno {\it et al.}. The agreement is again quite good (apart, perhaps, from the location of the minimum for $L=0$) and we conclude that our method of calculating these quantities is trustworthy. (We note that the converged ET-ERE calculations of $K_3$  are nearly identical with the FF results in Table \ref{Table1} and Figure~\ref{Fig2}.) Thus we can for the first time compare the HFD-B3-FCII for $K_3$ results with those employing other atom-atom potentials. As is apparent in Table~\ref {Table1}, the HFD-B result for $K_3(L=0)$ in the low energy limit is similar to the value for HFD-B3-FCII. The values for LM2M2 and TTY are similar to one another but larger than the other two by roughly a factor of five. The same behavior is also evident in Figure~\ref{Fig3} where FF results for $K_3(L=0)$ are plotted as a function of breakup energy. Nielsen and Macek~\cite{Nielsen:1999} as well as Esry, Greene and Burke~\cite{Esry:1999} found that the magnitude and shape of this quantity depends sensitively on  an interference of two distinct amplitudes. It is this interference which, for example, gives rise to the so-called Stueckelberg minima at $E_{breakup}=2000$-3000 $\mu$K. We speculate that, in the present work, the interference is between the purely two-body contributions and the three-body contribution appearing in Eq.~\ref{BigH}. In any case, all calculations display Stueckelberg minima which are qualitatively similar to that found originally in Ref.~\cite{Nielsen:1999} and later in Ref.~\cite{Suno:2002}. 

We next address extraction of universal functions which describe the recombination coefficients in the scaling limit~\cite{Braaten:2002jv,Braaten:2006vd} as outlined by Braaten, Kang, and Platter~\cite{Braaten:2006qx}. They find that, under a few physically reasonable simplifying assumptions, the recombination coefficient takes on the form (their Eq. 23)
\begin{equation}
K_3(L=0)=C_{max} |\sin[s_0 \ln({\bar a}_2/a_{0*})](1+h_1(x))+\cos[s_0 \ln({\bar a}_2/a_{0*})] h_2(x)|^2 \hbar {\bar a_2}^4/m
\label{KzeroUniversal}
\end{equation}
where the scaling variable is
\begin{equation}
x\equiv(m {\bar a}_2^2 E_{breakup}/\hbar^2)^{1/2},
\label{xUniversal}
\end{equation}
$C_{max}$ is a universal constant with a numerical value of about 402.7 and $m$ is the atomic mass. The quantity ${\bar a}_2$ depends on the dimer binding energy, $B_2$, via
\begin{equation}
{\bar a}_2=\sqrt{m B_2}/\hbar
\label{a2bar}
\end{equation}
while $a_{0*}$ is extracted from the value of $K_3(L=0)$ in the $E_{breakup}\rightarrow 0$ limit by solving Eq. 10 of Ref.~\cite{Braaten:2006qx}, namely
\begin{equation}
 K_3^{(0)}(L=0)=\frac{768 \pi^2(4\pi-3\sqrt{3})\sin^2[s_0 \ln({\bar a}_2/a_{0*})]}{\sinh^2(\pi s_0)+\cos^2[s_0 \ln({\bar a}_2/a_{0*})]}\ \frac{\hbar {\bar a}^4_2}{m}.
\label{astar}
\end{equation}
Values of ${\bar a}_2$ and ${\bar a}_2/a_{0*}$ for the four potentials we consider are given in Table~\ref{Table1}. 
Our values for $a_{0*}$ differ by less than 3$\%$ from those we obtain using the prescription of Ref.~\cite{Braaten:2006qx}, namely $a_{0*}\sim0.32\ \kappa_*^{-1}$ where
\begin{equation}
\kappa_*=\sqrt{2\mu_{1,2}\ B_{3*}^{(1)}}/\hbar
\label{kappastar}
\end{equation}
and where $\mu_{1,2}$ is the atom-dimer reduced mass while $B_{3*}^{(1)}$ is the ET-ERE LO trimer excited state binding energy in the limit that $a_2\rightarrow +\infty$. We note that a value of ${\bar a}_2/a_{0*}\simeq 1.15$ for HFD-B3-FCII is reported in Ref.~\cite{Braaten:2006qx} which is in accord with our result of 1.1477.

\begin{figure}[ht]
\vspace{0.50in}
\includegraphics[width=6in,angle=0]{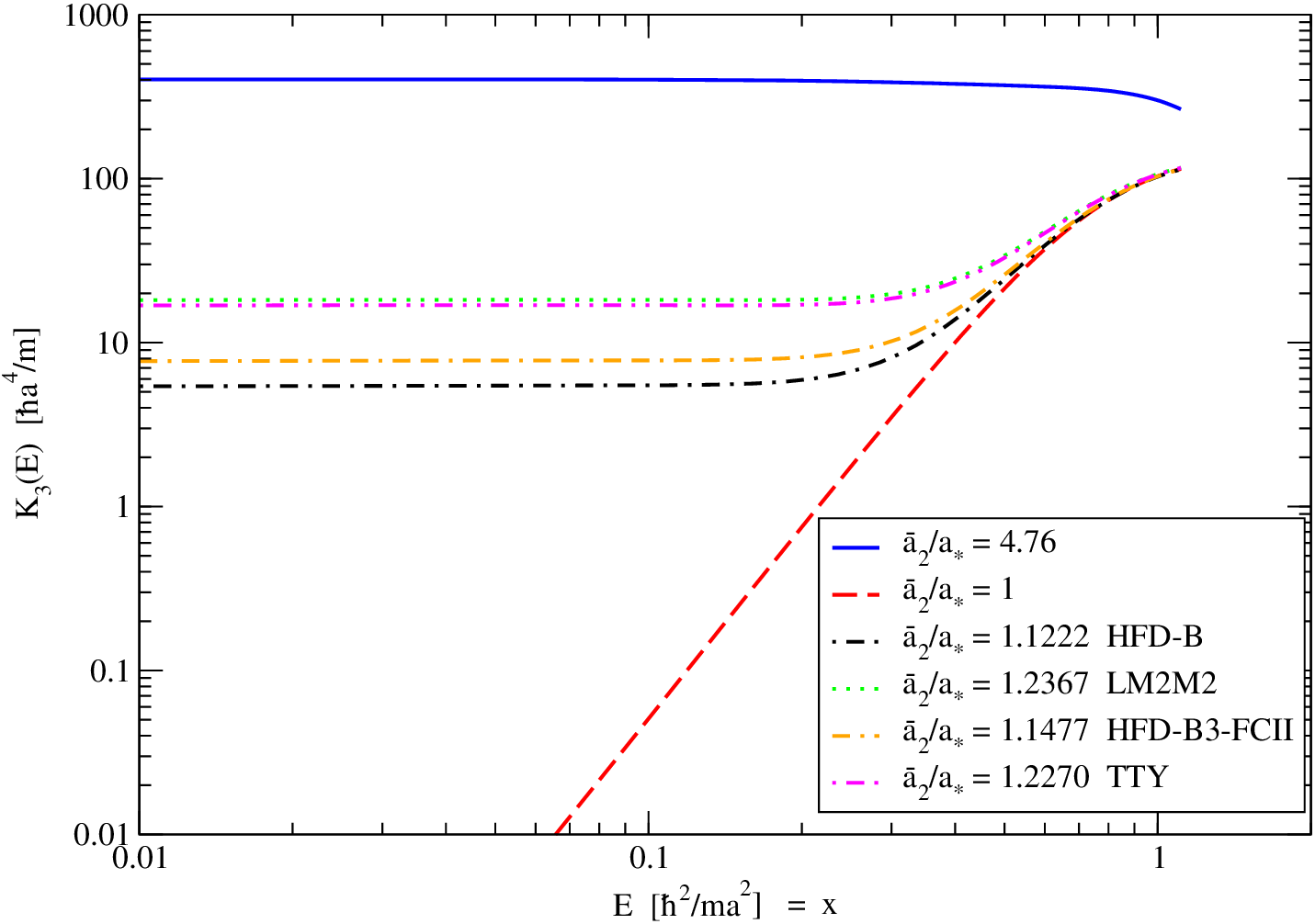}
\caption{(Color online).
         Sum of $K_3(L)$'s for $L\leq3$ computed using the universal functions,  $h_1(x)$ and $h_2(x)$, shown in Figure~\ref{Fig4} and various vales of ${\bar a}_2/a_{0*}$. These results are to be compared with Figure 2 of Ref. \cite{Braaten:2006qx}.
}
\label{Fig5}
\end{figure}


Because we have results for $K_3(L=0)$ for several atom-atom potentials, we can extend the work in Ref.~\cite{Braaten:2006qx} by solving Eq.~\ref{KzeroUniversal} for the two functions, $h_1(x)$ and $h_2(x)$, for the first time. We accomplish this by using the values of $K_3(L=0)$ obtained with potentials LM2M2 and HFD-B since these yield the highest and lowest values, respectively, of the $L=0$ recombination coefficient (see Table~\ref{Table1} and Figure~\ref{Fig3}). Our determinations of $h_1$ and $h_2$ are plotted in Figure~\ref{Fig4}. The range of the $h_i$'s extends up to $x\simeq 1.1$ which is just below the Stuekelberg minima where the assumptions used to obtain Eq.~\ref{KzeroUniversal} are no longer well-satisfied~\cite{Braaten:2006qx}. We have made polynomial fits to $h_1$ and $h_2$ and the results are also shown in Figure~\ref{Fig4}; the explicit forms of the fitted polynomials appear in the caption of Figure~\ref{Fig4}. We note that $h_1$ is much larger than $h_2$ 
and that both functions go to zero as $x$ goes to zero as should be expected from Eqs.~\ref{astar} and~\ref{KzeroUniversal}. 

Figure~\ref{Fig5} shows sums of $K_3(L)$'s for $L\leq 3$ in scaling units for a variety of values of ${\bar a}_2$ and $a_{0*}$. The results for ${\bar a}_2/a_{0*}=4.76$, 1 and 1.1477 can be compared directly with Figure 2 of Ref.~\cite{Braaten:2006qx} where only uncertainty bands appear. The present results are consistent with those of Ref.~\cite{Braaten:2006qx} except for the case of ${\bar a}_2/a_{0*}=1$ where our values fall below their narrow band. For this value of ${\bar a}_2/a_{0*}$, only $h_2(x)$ contributes (see Eq.~\ref{KzeroUniversal}) and, as mentioned above, we find $h_2(x)$ to be very small. It thus seems possible based on what appears in Figure~\ref{Fig5} that, for systems where ${\bar a}_2/a_{0*}\simeq 1$, $K_3$ could be two or three orders of magnitude smaller than the already small values found for potential HFD-B over a range of $E_{breakup}$ up to $100\ \mu$K (or $x\leq 0.2$). We also show in Figure~\ref{Fig5} calculations employing LM2M2 and TTY for which the values of ${\bar a}_2/a_{0*}$ and $K_3$ are relatively large. Dimensionful counterparts of the HFD-B3-FCII and TTY curves calculated using the universal functions are compared with the corresponding full calculations in Figure~\ref{Fig3}. Agreement is excellent and it is gratifying that the universal calculations can account for the dependence of $K_3(L=0)$ over the full range of variability caused by differences among the four atom-atom potentials we consider. (Note that the agreement for HFD-B and LM2M2 is exact by construction.)

\section{Conclusions}
\label{sec:conclusions}

We have performed calculations of three-body recombination rates for cold $^4$He using a new method which exploits the simple relationship between the imaginary part of the atom-dimer elastic scattering phase shift and the $S$-matrix for the recombination process. We compute the elastic phase shifts by solving three-body Faddeev equations in momentum space. These equations are relatively simple under our assumption of separable atom-atom interactions~\cite{Shepard:2007b}. When the form factors which characterize the separable interactions are taken to be $\delta$-functions, our elastic scattering calculations are identical to Effective Theory treatments based on the Effective Range Expansion (ET-ERE) at lowest order (LO) in the effective range expansion of the atom-atom potential. (See Ref.~\cite{Griesshammer:2004pe} for relevant discussion in the context of neutron-deuteron scattering and Ref.~\cite{Platter:2006ev} where $^4$He atom-dimer scattering is addressed.) We perform such calculations using four widely used atom-atom potentials, namely HFD-B3-FCII\cite{Aziz:1995}, HFD-B\cite{Aziz:1987}, LM2M2\cite{Aziz:1991} and TTY\cite{Tang:1995}. We also report ET-ERE results at next-to-leading order (NLO) and NNLO\cite{Griesshammer:2004pe} in the expansion parameter $r/a_2$ where $r$ is the range parameter in the effective range expansion of $p \cot\delta_0$ for atom-atom scattering. This quantity is quite small for $^4$He ($\sim 0.075$) and convergence is rapid. We also present results for a new kind of Faddeev calculation employing finite range form factors whose momentum dependence is fixed by fitting to $p \cot\delta_0$. These calculations have some similarities to neutron-deuteron scattering calculations done long ago, shortly after the Faddeev equations were formulated\cite{Aaron:1966,Watson:1967}. Recent work by Pen'kov~\cite{Penkov:2003} is also similar to ours. The ET-ERE calculations for the atom-dimer scattering lengths appear to be converging toward the FF results. Our converged ET-ERE results as well as our FF calculations of atom-dimer scattering lengths and phase shifts agree well with recently  published values by Motovilov {\it et al.}~\cite{Motovilov:1999iz}, by Roudnev~\cite{Roudnev:2002ab} and by Platter and Phillips~\cite{Platter:2006ev}. 

We have calculated three-body recombination coefficients, $K_3(L)$,  for $L\leq3$ using the four atom-atom interactions we consider. The HFD-B3-FCII results agree very well with the only previously published values of Suno {\it et al.}\cite{Suno:2002}. Our calculations using the other three potentials reveal that the recombinations coefficients vary by as much as a factor of six depending on these inputs. More specifically, the results for HFD-B3-FCII and HFD-B lie in the lower end of the range while LM2M2 and TTY values are nearly identical and establish the upper end. 

To the extent that the dynamics of the $^4$He atom-dimer system are governed by Efimov physics, it should be possible to describe them via scaling arguments~\cite{Braaten:2006vd,Braaten:2002jv}. In a recent work~\cite{Braaten:2006qx}, Braaten, Kang and Platter have shown that, under a few physically reasonable assumptions, the $L=0$ recombinations coefficient can be expressed in terms of ``universal'' quantities including two universal functions of a scaling variable. Because they only had access to the HFD-B3-FCII results of Suno {\it et al.}~\cite{Suno:2002}, they were unable to determine these functions with any precision. Having in hand recombination coefficients for four different potentials, we are able for the first time to determine these functions with considerable accuracy. Indeed, when they are fixed using the HFD-B and LM2M2 results, the universal relation describes the HFD-B3-FCII and TTY recombination coefficients very well. We also note that one of the universal functions we determine is quite small which suggests that  extremely small recombination coefficients can result if certain physical quantities are appropriately tuned.

\smallskip
\acknowledgments
The author would like give thanks to Chris Greene, Brett Esry and Jose D'Incao for invaluable discussions and suggestions and for help with preparation of the manuscript. Most of all, the author is grateful to these three for their patience and forbearance in the course of this project.  Finally, the author acknowledges numerous helpful conversations with Lucas Platter. This work was supported in part by DOE grant DE-FG05-92ER40750.

\vfill\eject
\bibliography{../ReferencesJRS}

\end{document}